\documentclass{PoS}
\usepackage[utf8]{inputenc}
\usepackage{amsmath}
\usepackage{graphicx,subfigure}
\usepackage{float}
\usepackage{natbib}
\bibpunct{(}{)}{;}{a}{}{,}

\title{RFI mitigation for the Effelsberg-Bonn HI Survey (EBHIS)}

\ShortTitle{RFI mitigation for the Effelsberg-Bonn HI Survey (EBHIS)}

\author{\speaker{Lars Flöer}\\
       Argelander-Institut für Astronomie (AIfA)\\
       E-mail: \email{lfloeer@astro.uni-bonn.de}}

\author{Benjamin Winkel\\
       Argelander-Institut für Astronomie (AIfA)\\
       E-mail: \email{bwinkel@astro.uni-bonn.de}}

\author{J\"urgen Kerp\thanks{PI}\\
       Argelander-Institut für Astronomie (AIfA)\\
       E-mail: \email{jkerp@astro.uni-bonn.de}}

\FullConference{RFI mitigation workshop\\
		 29-31 March 2010\\
		 Groningen, the Netherlands}
		 
\abstract{A new L-band 7 feed array at the 100\,m telescope is used to perform an unbiased, fully sampled HI survey of the whole northern hemisphere -- the Effelsberg-Bonn HI Survey (EBHIS). The use of state-of-the-art digital Fast Fourier Transform spectrometers based on FPGAs -- superior in dynamic range and allowing fast dumping of spectra -- makes it possible to apply sophisticated RFI mitigation schemes.

Based on the current status of the survey we discuss the RFI situation at the 100\,m telescope and present a fast algorithm to automatically identify RFI in the raw data output from the spectrometer. Using simulations we show that it is feasible to detect more than 95\% of all RFI in excess of 1$\sigma$ amplitude with less than 1\% false positives.}

\begin{document}

\section{The Effelsberg-Bonn HI Survey}

The Effelsberg-Bonn HI Survey (EBHIS) is a new all-sky HI survey of the northern hemisphere for both galactic and extragalactic emission \citep{2008glv..book..345W}. It is conducted with the 100\,m Effelsberg telescope using a 7 feed array frontend and modern field programmable gate array (FPGA) implemented fast fourier transform (FFT) spectrometers.

The 7 feed array was originally built for RADAR-experiments targeting space debris and is highly resistant to saturation from strong signals. The beam size is 9' with a separation of 15' between the individual beams. Measurements are carried out in on-the-fly mode in equatorial coordinates with a typical mapsize of 25 square degrees. By rotating the array with 14$^\circ$ with respect to the scanning direction we obtain 7 equidistant scanlines with a separation of 4' which yields a fully sampled grid.

The spectrometer provides a bandwidth of 100\,MHz with 16384 channels. This results in an effective velocity resolution of $\Delta v = 1.45$\,km/s and allows to cover the local standard of rest (LSR) velocity range from $-1,000$\,km/s to $+ 20,000$\,km/s simultaneously. We therefore map the galactic and extragalactic HI emission in one pass. The backend also allows for fast dumping of spectra. We chose a compromise between data rate and time resolution to apply RFI mitigation and dump spectra every 500\,ms. This results in a typical data volume of 5\,GB per 25 sq. deg. measurement.

For further details about the setup and data reduction process we refer to \citet{Winkel:2010fk}.

\section{RFI in EBHIS data}

\begin{figure}[]
\begin{center}
\includegraphics{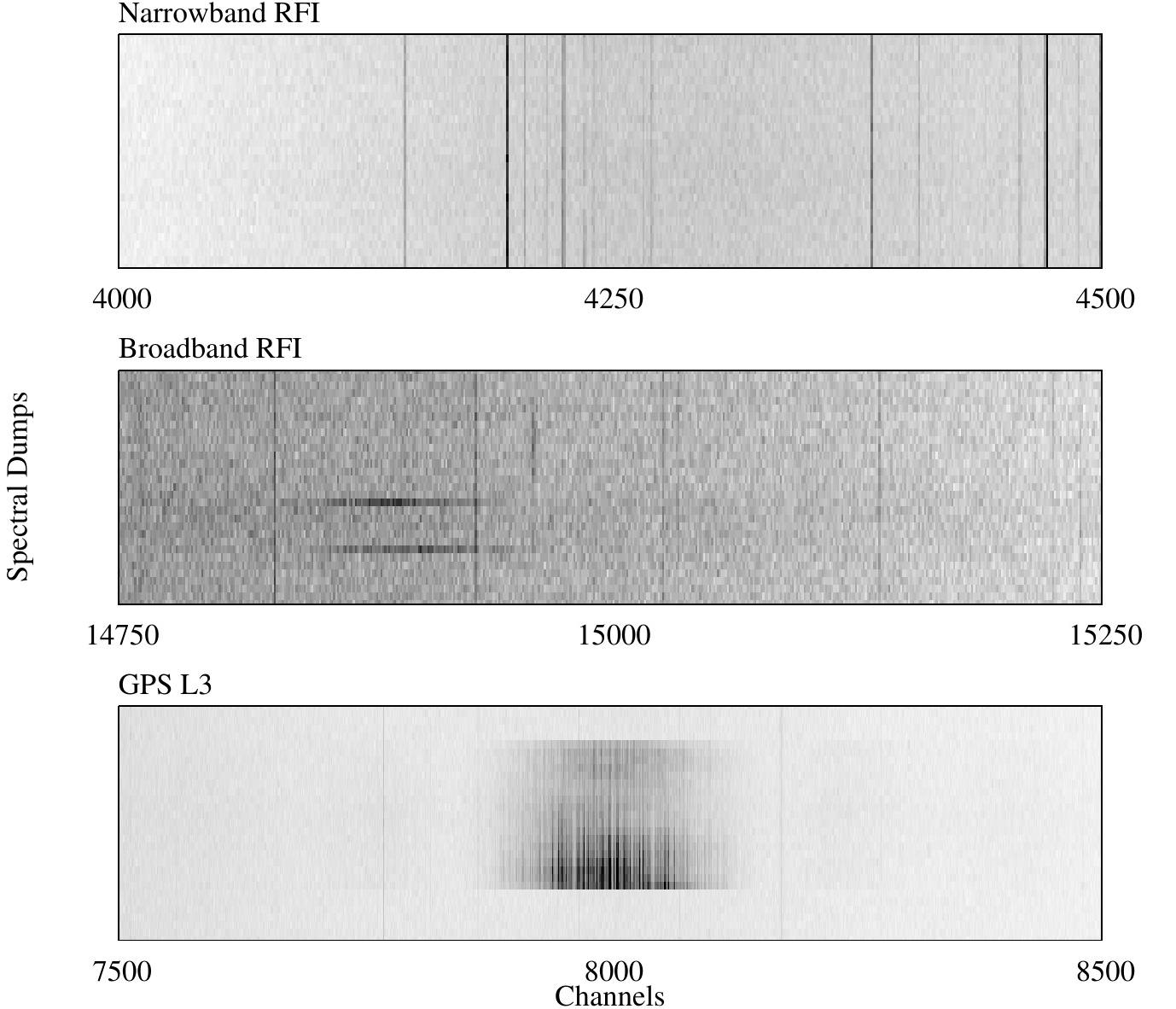}
\caption{Examples of RFI in EBHIS data. Each panel contains 30 spectral dumps. The top panel shows typical narrowband RFI. The middle panel presents an example of two broadband RFI events. The bottom panel shows the GPS L3 mode.}
\label{fig:examplerfi}
\end{center}
\end{figure}

Analysis of the EBHIS raw data shows two main types of RFI. The bulk of RFI are narrowband signals, contaminating a single or two adjacent channels. More than 95\% of the narrowband RFI are constant in amplitude and even the varying RFI are constantly detectable. The other, less common, type of RFI are short, broadband events with a characteristic bandwidth of order of 100 channels ($\simeq 600$\,kHz) and durations of $\tau \leq 500$\,ms, since they are only present in a single dump. The spectral shape of broadband RFI can be described by a Gaussian. 

There are other types of RFI much less common, that do not fall in one of either before mentioned categories. An example of such RFI is the GPS L3 mode at 1381\,MHz which corresponds to center frequency of our observing band. Examples of the RFI described here are shown in Fig.\,\ref{fig:examplerfi}.

\begin{figure}[]
\begin{center}
\includegraphics{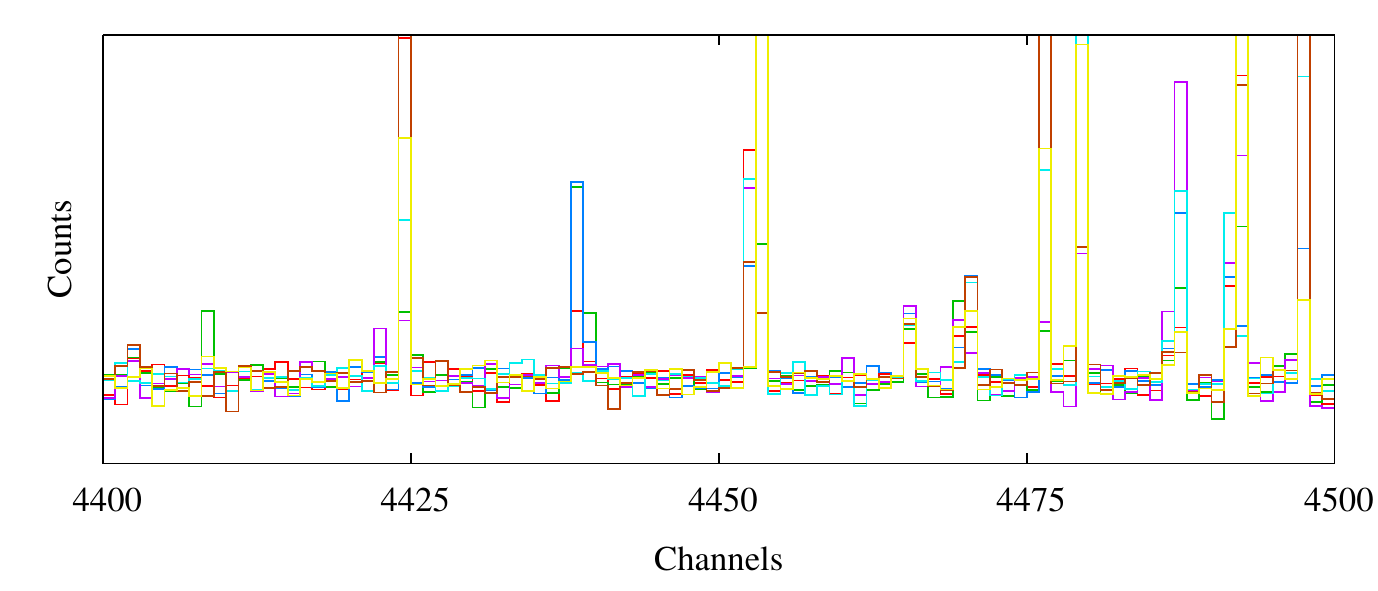}
\caption{Example of narrowband RFI affecting the 7 horns with different intensity.}
\label{fig:basebandcompare}
\end{center}
\end{figure}

We also observe that all horns of the 7 feed array are affected similarly by external RFI. An example of the RFI contamination in the 7 different feeds can be seen in Fig.\,\ref{fig:basebandcompare}. The narrowband RFI is visible in all spectra and only varies in amplitude across the feeds. This is due to different paths and different polarizations measured by the horns. This fact will be exploited for detection of very weak RFI signals in the spectra.

\section{An algorithm for automated RFI detection}

To automatically detect RFI in our data we make use of the fact, that both main types of RFI --- narrowband and broadband --- are very well defined features in either frequency or time. This distinguishes them from other signals like galactic HI emission or galaxies which are both smooth features. To extract the RFI we use an approach based on cross-correlation
\begin{equation}
C(x) = \sum_{i = -n/2}^{+n/2} (p(i) - \bar{p})\cdot(s(x+i) - \bar{s})
\end{equation}
with templates $p$ of size $n$
\begin{equation}
p(i) = \left\{
\begin{array}{lr}
    1&:  i = 0    \\
    0&:  i \neq 0   
\end{array}
\right.
\end{equation}
that match the shape of the RFI, either in time or frequency, best. Here, $\bar s$ is the mean of the spectral data covered by the template, $\bar p$ is the mean value of the template and $s(x)$ and $p(x)$ are the spectral data and value of the template at $x$, respectively. In practice, strong RFI has a large impact on $\bar s$ and the product $\bar p \cdot s(x)$ can produce troughs in the correlation spectrum. To avoid this effect, we set $\bar p = 0$ and replace $\bar s$ by the median estimator. With these alterations, the cross-correlation reduces to the subtraction of a running median and it has been implemented in that way. In general, this approach also allows for custom filters, e.g. to extract the shape of the GPS L3 mode signal from the data.

Furthermore, the algorithm makes use of the fact that for any given dump we have 14 spectra (7 horns times 2 polarizations) that are all affected similarly by external RFI. A challenge in automated RFI mitigation is to detect low signal-to-noise RFI in the data. For faint RFI, individual thresholding of single spectra at a certain level $t_{1}$ might not be sufficient. We circumvent this shortcoming of  thresholding by requiring that a certain number $N$ of simultaneous values all lie above a threshold $t_{N} < t_{1}$. The more values one requires, the lower the threshold. Assuming Gaussian noise with standard deviation $\sigma$, the threshold for $N$ values $t_{N}$ that has the same theoretical false detection rate as $t_{1}$ for a single value can be estimated from
\begin{equation}
1 - {\rm erf}\left(\frac{t_{1}}{\sqrt{2}}\right) = \left[1 - {\rm erf}\left(\frac{t_{N}}{\sqrt{2}}\right)\right]^{N}\quad.
\end{equation}
Here erf is the error function of the normal distribution and $1 - {\rm erf}(x/\sqrt{2})$ gives the probabilty that a Gaussian distributed random variable is not contained in the interval $[-x\sigma:x\sigma]$.

As an example, if one decides to flag all $t_{1}  = 5\sigma$ outliers in a single spectrum the corresponding threshold for $N = 4$ spectra would be $t_{4} \approx 2.2\sigma$ and for $N = 14$ spectra it would be as low as $t_{14} \approx 0.97\sigma$.

The algorithm then works as follows:
\begin{enumerate}
\item Subtract a windowed median from each spectrum.
\item Calculate the sample standard deviation $\sigma$ for each spectrum.
\item Take $N$ spectra and flag values that are above $t_{N}$ in all $N$ spectra. Only positive outliers are flagged, since RFI is an additive effect.
\item Recalculate the sample standard deviation $\sigma$ for each spectrum and drop the already flagged values in the process. This is done because heavy RFI strongly affects the sample standard deviation.
\item Repeat 3. and 4. until no additional flags are found.
\end{enumerate}

This scheme is applied for all $N$ between 1 and 14 in increasing order.

In practice we apply this algorithm in frequency and time. To flag the narrowband RFI we average over the dumps of one scan line and pass the average spectrum to the algorithm. For the broadband RFI we perform a 16 fold binning in the spectral domain and apply the algorithm to the time evolution of each bin. The averaging and binning is performed to increase the signal-to-noise of the RFI and in case of the broadband RFI to detect also the ``wings'' of the RFI event close to the noise level. Both flagging stages use their own starting threshold $t_{1}$ and we denote the threshold for narrowband flagging with $t^{n}_{1}$ and for broadband flagging with $t^{b}_{1}$.

Before the algorithm can safely be applied, we prepare the spectra by subtracting the continuum level to overcome the continuum radiation sources and elevation effects, as both could lead to false broadband RFI detections. Furthermore, strong outliers are flagged beforehand by subtracting the median from the mean of a scan line. Signals in excess of $5\sigma$, like the GPS L3 mode, are then marked as bad data.

After the successful flagging of the data, the flags are stored in a MySQL\footnote{http://www.mysql.org} database for later use. This ensures accurate bookkeeping and also gives the possibility for a statistical analysis of the RFI encountered, e.g. RFI amplitude as a function of azimuth and elevation of the telescope.

\section{Simulations}

To quantify the performance and optimize the thresholds of the used algorithm we carried out simulations.

\subsection{Simulated data}

\begin{figure}[]
\begin{center}
\includegraphics{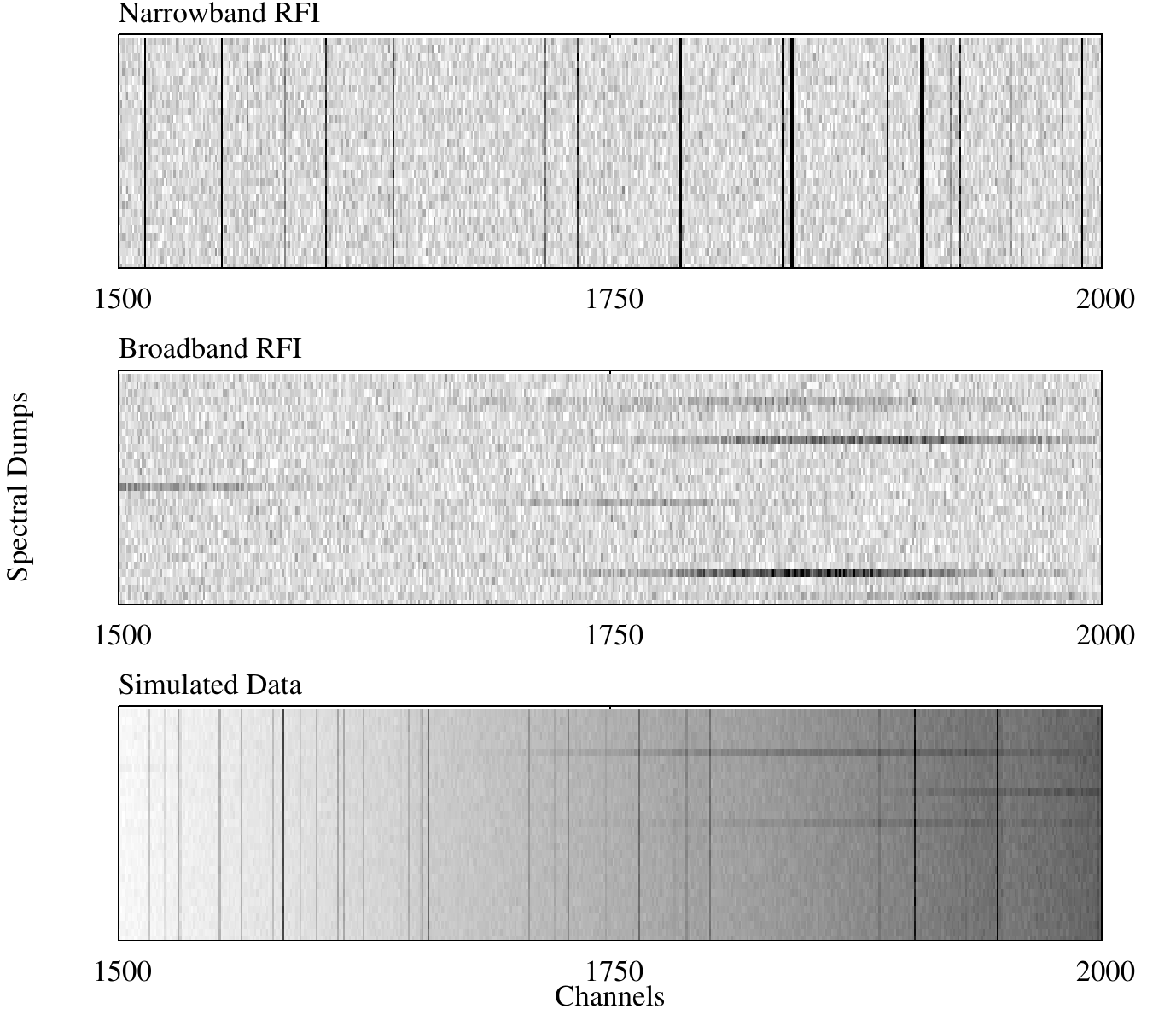}
\caption{Examples of the datasets generated. Top panel: Narrowband RFI and noise. Middle panel: Broadband RFI and noise. Bottom panel: Narrowband and Broadband RFI incorporating baseline effects.}
\label{fig:simsets}
\end{center}
\end{figure}

We generated 14 sets of 30 spectra with 2048 channels to mimic the amount of real data obtained in one scan line. Each spectrum contains Gaussian noise with standard deviation $\sigma$. Based on an earlier RFI census at Effelsberg by \citet{2007AN....328...68W}, we have drawn the RFI amplitudes from the tail of a Rayleigh distribution with lower limit of $0.25\sigma$. To simulate that the RFI are present in all 14 simultaneous spectra, we chose a uniformly distributed, random fraction between 0 and 1 of the generated RFI flux and inserted them in the same, randomly chosen channel in each spectrum. To simulate the extended characteristic of broadband RFI we multiplied the generated RFI flux with a gaussian of random width and cut off at when they got below $0.25\sigma$ amplitude.

We generated three different types of datasets:
\begin{enumerate}
\item Gaussian noise and narrowband RFI
\item Gaussian noise and broadband RFI
\item Gaussian noise, narrowband and broadband RFI and a fifth-order polynomial with added sine component of random phase to simulate baseline effects.
\end{enumerate}
Each type of dataset was generated 100 times and then passed to the flagging algorithm. In the results we show the mean and the standard deviation of all computed values.

\subsection{Results}

For each simulation we generated a histogram, showing the average abundance of RFI with a certain intensity and their detection. Those plots also show a cumulative detection rate as a function of RFI amplitude, which is the fraction of the total RFI contamination that we are able to detect down to that particular level.

We also generated accuracy plots, that show the performance of the individual flagging step for different starting thresholds $t_{1}$. Here we plot the fraction of RFI detected as a function of the fraction of flags that are wrongly set. This is not to be confused with the fraction of data lost, which is much lower. In these plots we differentiate between the total RFI count and RFI with amplitudes in excess of $1\sigma$. These plots are comparable to the ones found in \citet{2010MNRAS.tmp..410O}, where the authors compared the performance of different mitigation schemes.

\begin{figure}[]
\begin{center}
\includegraphics{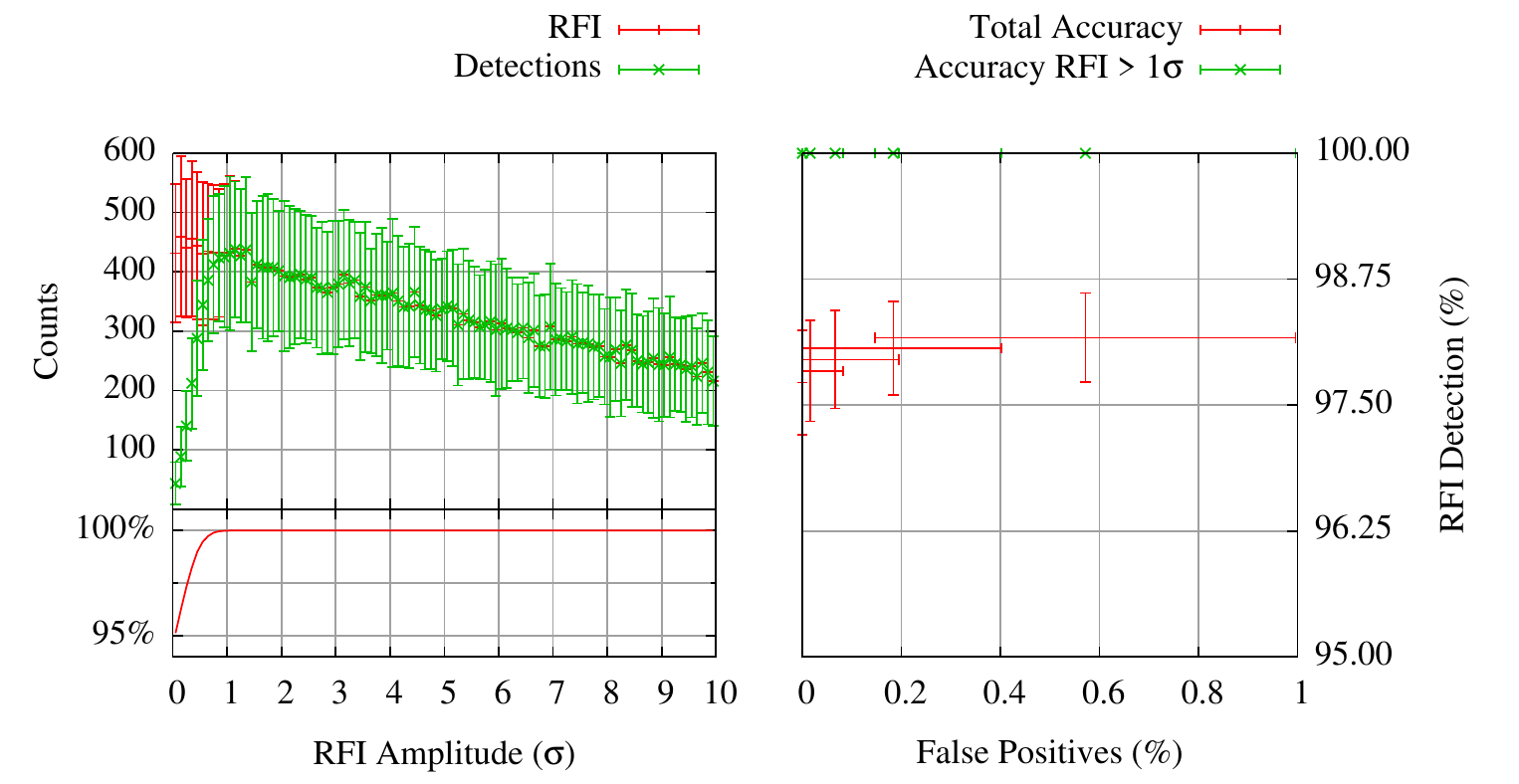}
\caption{Results from the narrowband RFI simulation. The left hand plot shows the average RFI distribution and their detection for $t^{n}_{1} = 7\sigma$. The bottom of the plot shows the cumulative detection fraction. The average false positives ratio for this case is about 0.1\%. The right hand side plot shows the fraction of RFI detected in dependence of the fraction of flags that are false positives for different thresholds $t^{n}_{1}$. This gives a measure of the accuracy of the algorithm.}
\label{fig:nbresults}
\end{center}
\end{figure}

Figure \ref{fig:nbresults} shows the results for the narrowband RFI test case. We are successful to detect more than 97.5\% of all narrowband RFI with less than 0.1\% false positives. If one only considers RFI in excess of $1\sigma$ the detection rate reaches 100\%. 
\begin{figure}[]
\begin{center}
\includegraphics{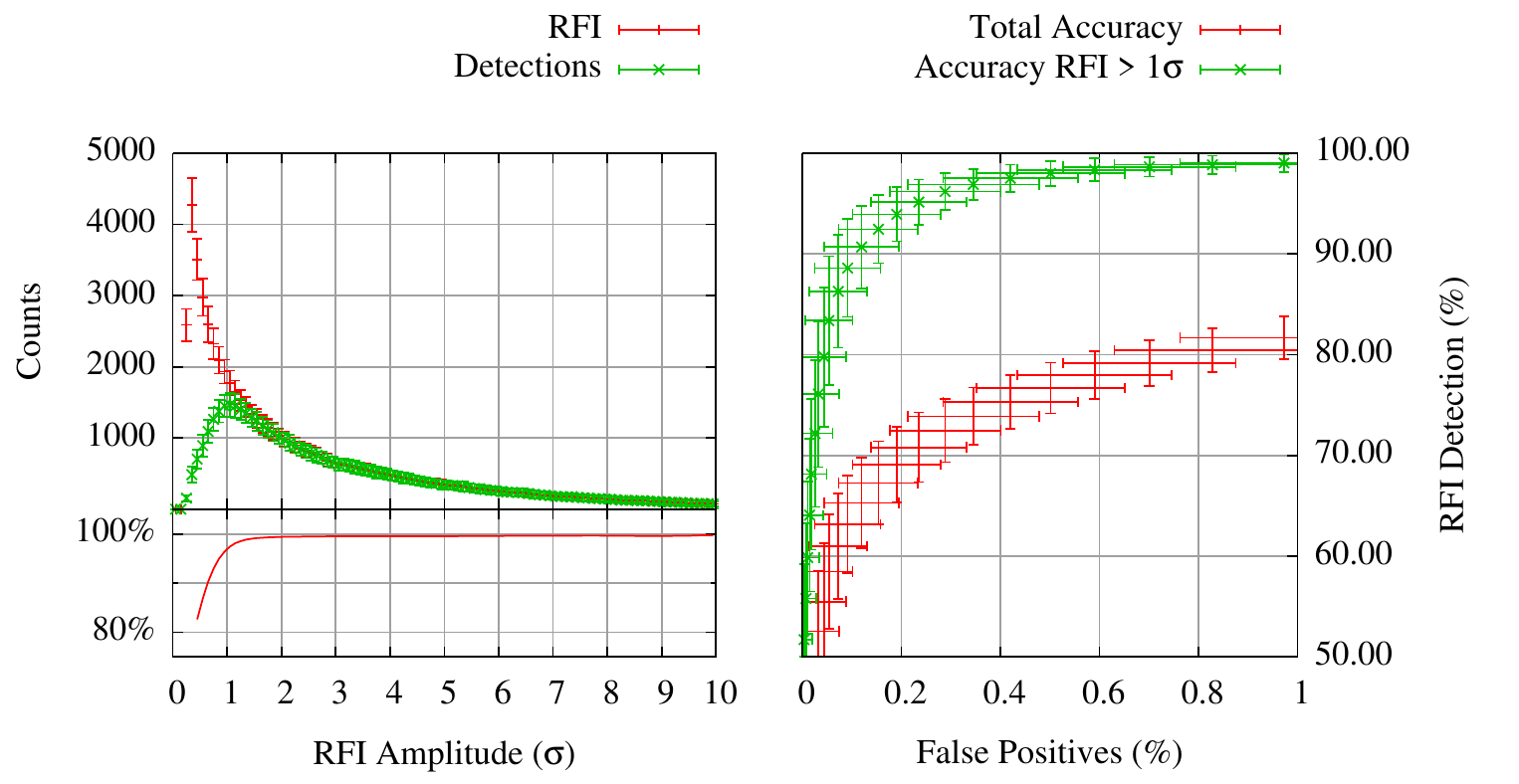}
\caption{Same viewgraphs as in Fig.\,{\protect\ref{fig:nbresults}} for broadband RFI. The histogram shows the results for $t^{b}_{1} = 7\sigma$ with a false positives ratio of 0.5\%.}
\label{fig:bbresults}
\end{center}
\end{figure}
Figure \ref{fig:bbresults} shows the same viewgraphs for the broadband RFI tests. Here we detect about 80\% of all RFI with a false positves fraction of about 1\%. If one only considers again RFI in excess of $1\sigma$ the detection rate is in excess of 95\% with less than 0.5\% false detections.

\begin{figure}[]
\begin{center}
\includegraphics{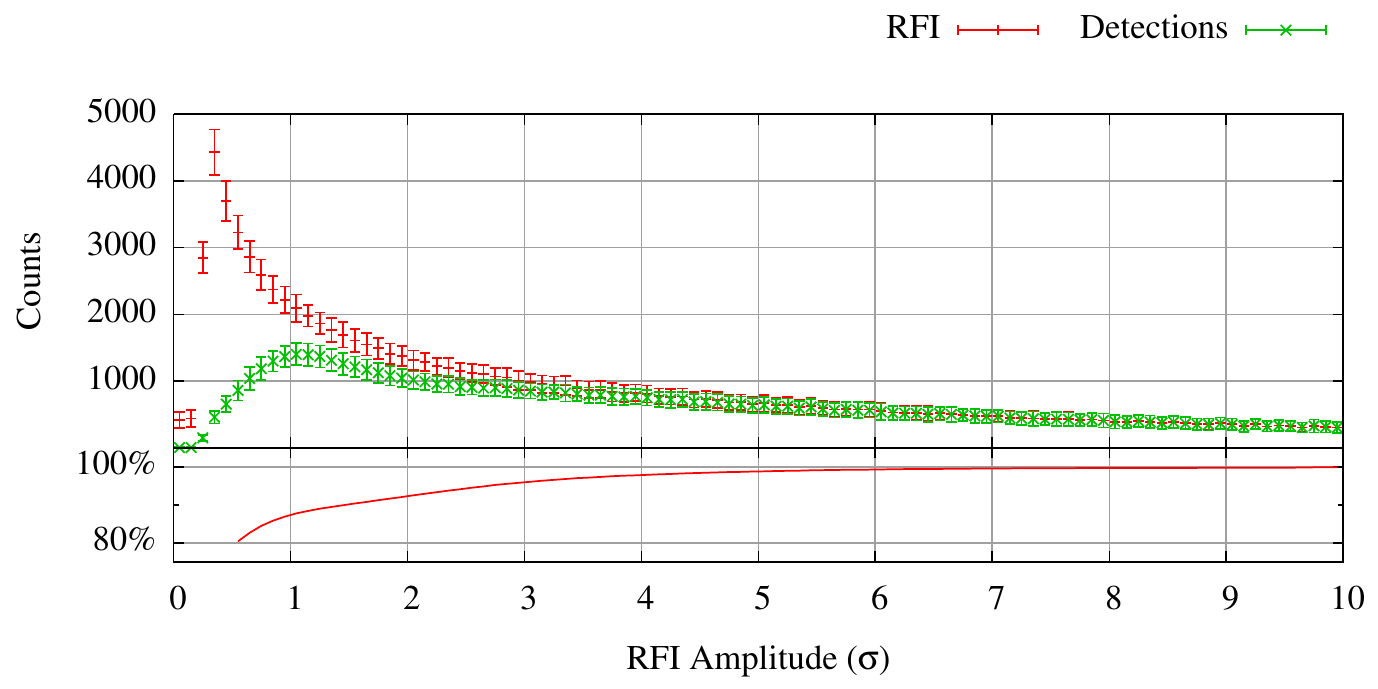}
\caption{Results from the complete RFI simulation. The plot shows the average RFI distribution and their detection for $t^{n}_{1} = 7\sigma$ and $t^{b}_{1} = 7\sigma$. The bottom of the plot shows the cumulative detection fraction. The average false positives ratio for this simulation is about 0.3\%.}
\label{fig:normalresults}
\end{center}
\end{figure}

The results for the more realistic test case are shown in Fig.\,\ref{fig:normalresults}. The histogram shows the detection rate for a parameter set of $t^{n}_{1} = 7\sigma$ and $t^{b}_{1} = 7\sigma$. It shows that we are able to detect 90\% of all RFI in excess of $1\sigma$. The false positives ratio for this histogram is about 0.3\%. If one tolerates a false detection rate of about 1\% this detection ratio increases to about 95\%. We account this decrease of performance to the inaccuracies introduced by the simulated baseline, though we expect the algorithm to perform better on real data, since the real baseline is much smoother.

Furthermore, it has to be mentioned that he abundance of broadband RFI is highly exaggerated in the simulations. In real data, broadband RFI is sparse and only prominent in the low frequency part of our observing band. We therefore expect a lower fraction of false positives and higher fraction of true positives for real data.

\section{First Results}

\begin{figure}[htbp]
   \centering
   \includegraphics[width=0.49\textwidth]{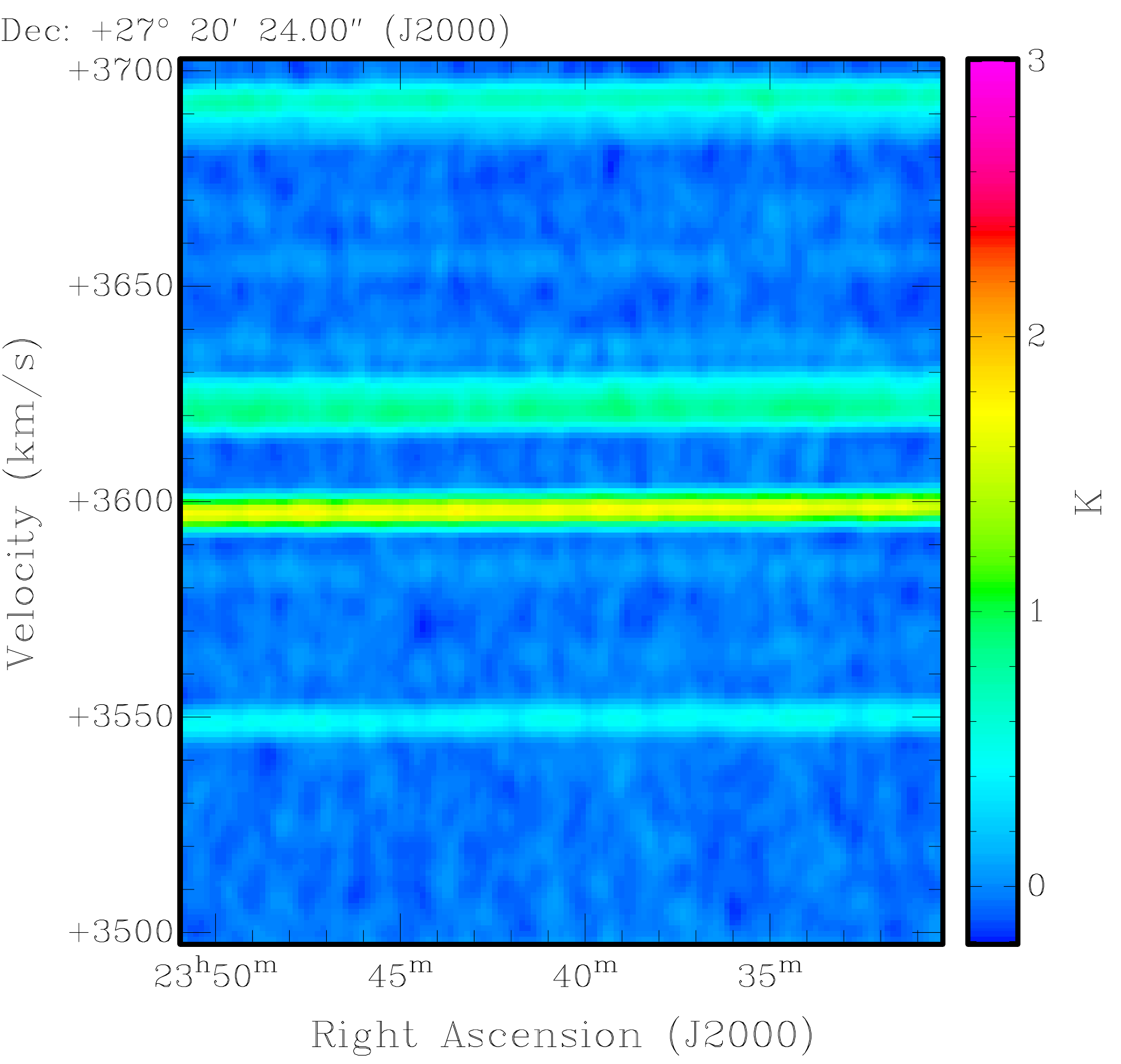}
   \includegraphics[width=0.49\textwidth]{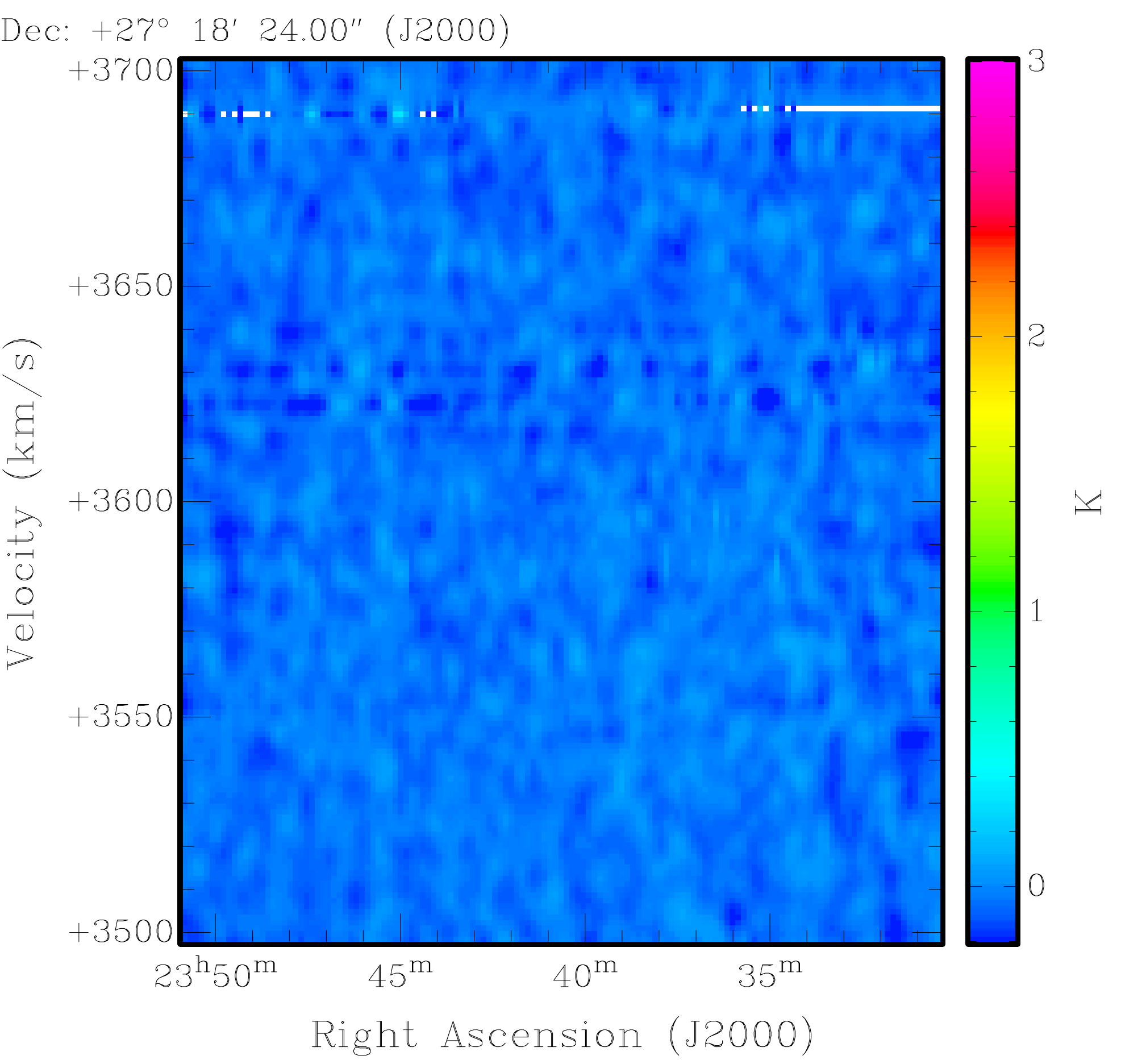}
   \caption{A position-velocity ($\alpha$ and $v_{\rm LSR}$) plot of EBHIS data. Left panel: Original data with RFI contamination. Right panel: The same data with the flagged values blanked. The locally increased noise is visible especially between 3600\,km/s and 3650\,km/s. The total noise of the RFI contaminated data shown is 321\,mK whereas the noise of the cleaned data is 50\,mK, which matches the theoretical expectation.}
   \label{fig:example}
\end{figure}

The algorithm has already been applied to various real data sets in two different ways. The most simple application of the obtained flags is the blanking of the flagged values and gridding the final data cube without those values. One can clearly see an improvement in data quality but also locally increased noise or even blank pixels where there are no values left. An example of a processed datacube can be seen in Fig.\,\ref{fig:example}. The flagged values have been blanked and the datacube was smoothed using a Hanning window.

Another approach has recently been tested by the subtraction of the narrowband RFI that are constant in intensity. For those, we subtract the median of the RFI flux and interpolate the RFI-free flux of the channel from the adjacent ones. This simple approach works very well in increasing data quality and does not introduce higher noise levels or missing values.

\section{Conclusions}

The algorithm described takes advantage of the the RFI contamination in the different feeds of the 7 feed array to detect RFI below the noise level of the raw data. For RFI in excess of $1\sigma$, simulations have shown a detection rate of over 95\% for simple and about 90\% for more realistic cases. In both cases, the false detection ratio is below 0.5\%. First applications to real data show great improvement in data quality.

\acknowledgments
Based on observations with the 100-m telescope of the MPIfR (Max-Planck-Institut für Radioastronomie) at Effelsberg.

\bibliographystyle{aa}
\bibliography{mititgation}

\end{document}